# Compute at Scale
# A Broad Investigation into the Data Center Industry

**Konstantin Pilz[1] and Lennart Heim**

*July 2023*

## Abstract

This report characterizes the data center industry and its importance for AI development. Data centers are industrial facilities that efficiently provide compute at scale and thus constitute the engine rooms of today's digital economy. As large-scale AI training and inference become increasingly computationally expensive, they are dominantly executed from this designated infrastructure. Key features of data centers include large-scale compute clusters that require extensive cooling and consume large amounts of power, the need for fast connectivity both within the data center and to the internet, and an emphasis on security and reliability. The global industry is valued at approximately $250B and is expected to double over the next seven years. There are likely about 500 large (> 10 MW) data centers globally, with the US, Europe, and China constituting the most important markets. The report further covers important actors, business models, main inputs, and typical locations of data centers.

---

[1] Correspondence to mail@konstantinpilz.com

## Summary

| Metric | Estimate |
| --- | --- |
| Number of data centers globally | 10,000 - 30,000 (70% confidence interval) |
| Number of large (>10 MW) data centers globally (size that could host a large supercomputer[2]) | 335 - 1,325 (70% confidence interval) |
| Typical power capacity of a newly constructed data center | ~20 MW (could host a cluster of about 16,000 NVIDIA H100 GPUs[3]) |
| Total power consumption of the industry | ~45 GW; 1 - 2% of global electricity consumption |
| Construction cost of a typical data center (20 MW), excluding hardware | ~$100 - 200M |
| Data center market size | ~$250B |
| Market growth | ~10% /year → doubling every seven years |
| Average size of a data center campus | ~10,000 sqm (1.5 football pitches) |

*Table 1: Estimates of key metrics and links to more thorough coverage.*

This report provides an overview of the data center industry and its relationship with AI development. It primarily reviews existing literature, occasionally drawing conclusions for AI governance *(See Scope)*. A subsequent piece comments on the role data centers may play in mitigating risks from advanced AI systems.

**Data centers are purpose-built industrial facilities that host hardware at scale and thus efficiently provide computational resources (compute).** They primarily run various internet services required for banking, web browsing, online gaming, communications, video streaming, etc. Besides, **some data centers host high-performance compute clusters to run computationally intensive operations** such as scientific simulations and **machine learning (ML).** Data centers form an integral **part of the (AI) compute supply chain** constituting the link between the semiconductor industry and the compute end-users. Right now, the reader likely interacts with several data centers—by accessing this document, receiving text messages, synchronizing files, or updating newsfeeds. *(More in section What are data centers?.)*

**Key features** of a large data center include:
- Tens to hundreds of MW of **power consumption**, similar to that of a medium-sized city (~100,000 inhabitants).
- Extensive **heat production** requiring immense cooling systems consuming water and additional power.

---

[2] Though only a small fraction of all data centers host supercomputers.
[3] The DGX H100 has a power capacity of 10.2 kW and contains 8 H100. 20,000 kW / 10.2 kW * 8 = 15,686. (This is a simplified estimate.)

- Emphasis on **redundant components and backup systems** such as power generators to ensure high reliability.
- **Physical security measures** to prevent unauthorized access.
- **Spatial requirements** similar to other industrial facilities, with 10,000 to 100,000 sqm, the equivalent of several football pitches.
- High-speed data transmission, requiring **low latency, high bandwidth connections** both within and from/to the data center.
- **Complex supply chain** management due to a high number of specialized inputs.

*(More in section [Key characteristics](Key characteristics).)*

Today, AI development increasingly depends on data centers as **training large ML systems requires dedicated compute clusters of thousands of high-bandwidth interconnected AI accelerators[4] that are operated from large data centers.** Further, **the efficient deployment of ML models on a large scale** (e.g., offering ChatGPT as a service) **similarly requires this designated infrastructure**. Understanding the data center industry hence sheds light on the global distribution of (ML) compute and which actors can, in principle, train large ML systems. The industry's potential for growth also determines how quickly AI technology can be adopted widely. Further, data centers may present an opportunity for monitoring and regulating AI development and deployment. *(More in sections [Data center's relevance for AI governance](Data center's relevance for AI governance), and [Shift to the cloud](Shift to the cloud).)*

Data centers can roughly be divided into:
- (i) 60% self-owned, "**enterprise**" data centers, where the hardware user owns and operates both the hardware and the infrastructure, and
- (ii) 40% shared "**colocation**" data centers, where a specialized company owns and operates the infrastructure (power, cooling, connectivity, security, backup systems) to host the hardware of other entities.
- For both configurations, the hardware hosted by the data center can be used directly by its owner, called **on-premises**, or to provide **cloud compute** that is rented out online, called **off-premises**. *(More in section [Types of data centers](Types of data centers). See [Figure 6](Figure 6) for a visual overview.)*

There are **an estimated 110 - 225 of the largest data centers**[5] with a power capacity[6] of above 100 MW and **225 - 1,100 large data centers** with a capacity of 10-100 MW, the size that could currently host an AI compute cluster for a major training run[7]. These large data centers are predominantly constructed by tech giants such as Google, Amazon, Microsoft, Meta, and Apple. **Including smaller builds, starting at 0.1 MW, there are 10,000 - 30,000 data centers globally**. Although data is limited, roughly **a third of them are likely in the US, followed by 25% in Europe and 20% in China**. While most data centers are close to major cities to allow for low-latency connections, large data centers are increasingly constructed in more remote places due to their spatial and power requirements. *(More in sections [How many data centers are there?](How many data centers are there?), [Locations of data centers](Locations of data centers), and [Most important companies](Most important companies).)*

The data center **market is valued at about $250B and projected to more than double in the next seven years**[8]. The colocation sector is currently shared by more than a dozen smaller actors, with the biggest company, Equinix, accounting for 11%; however, it appears to be slowly getting more

---

[4] AI accelerators refer to different types of chips specialized for AI applications such as GPUs and TPUs.
[5] All ranges are 70% confidence intervals.
[6] Including all power consumption (hardware and infrastructure).
[7] Though only a small fraction of such data centers likely host considerable AI compute clusters.
[8] This excludes the semiconductor market.

concentrated. Meanwhile, the **cloud market[9] is already dominated** by Amazon Web Services (**AWS**) (34%), Microsoft **Azure** (21%), and **Google Cloud** (11%). Due to economies of scale, ML applications are increasingly run on cloud services, leading to significant compute aggregations at cloud companies. *(More in sections [Market size and growth](), [Most important companies](), and [Shift to the cloud]().)*

A **typical investment into the supporting infrastructure for a 20 MW data center is around $100 - 200M** (excluding hardware)**,** and a large data center campus (> 100 MW) can cost up to a billion dollars. The costs are due to the requirements of **specialized equipment** such as cooling infrastructure, power transformers, high-speed connectivity components, and backup systems. (*More in section [Key Inputs for data center construction]().*)

A simple estimate suggests that **operating a large data center costs at least single-digit millions per MW per year**. This is mainly due to the large quantities of power it consumes, but also due to the expensive maintenance of computer hardware and other components. (*More in section [Key inputs for data center operation]().*)

Even in scenarios of explosive demand for AI, global data center capacity could unlikely grow by more than 40% per year. This is because the high number of specialized inputs needed for construction leads to bottlenecks in the supply chain, as has recently happened during the COVID-19 pandemic. Furthermore, spare power grid capacity is already limiting growth in several regions, and large cloud providers struggle to find suitable sites for their large data centers. Additionally, technical limits in power consumption and heat dissipation could make future compute clusters increasingly expensive, even barring fast growth. (*More in section [Limiting factors for data center growth]().*)

---

[9] Market analysts separate the data center market, including infrastructure inputs and services on the data center level, and the cloud market, including the services cloud providers offer.

## Visual overview

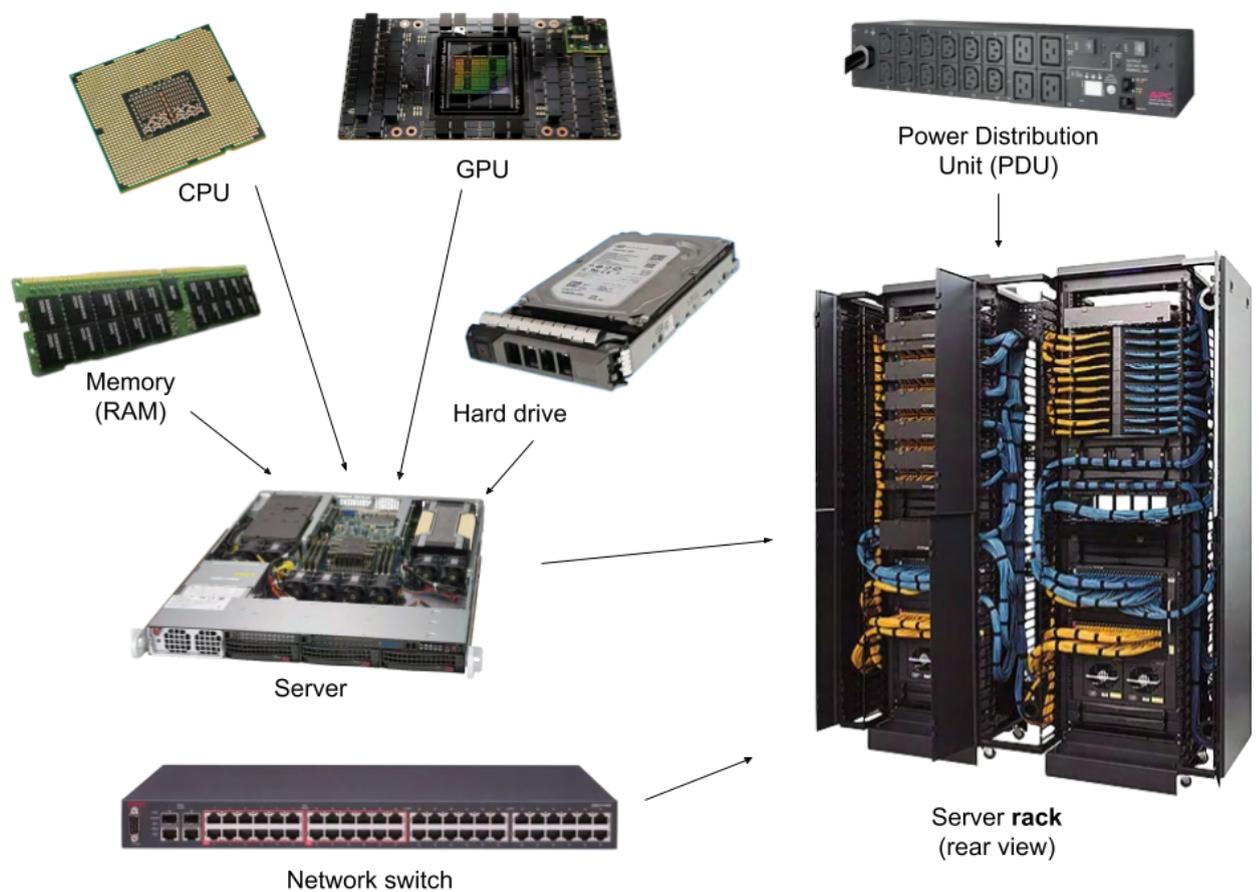

*Figure 1 - Selected computer hardware components in a typical data center, not to scale.[10] The **CPU** (Central Processing Unit) is a general-purpose processor, while the **GPU** (graphics processing unit) is a specialized chip for graphics rendering and machine learning. (There are several other chips specialized for AI training, such as Google's Tensor Processing Unit (TPU).) Random Access **Memory** (RAM) temporarily stores data that can quickly be accessed, while **hard drives** store data long-term. In addition to other components, these parts comprise a **server**, the smallest functional unit in a data center, which is approximately analogous to a desktop computer without any input or output devices. Several servers are arranged into a **rack**, where they receive power via a **Power Distribution Unit** (PDU) and a connection to other servers and the internet via a **network switch**. (More in section [Key characteristics](#).)*

---

[10] This is a general-purpose setup. Servers for AI specific workloads contain a higher number of GPUs and more RAM. See e.g., [NVIDIA's DGX](#).

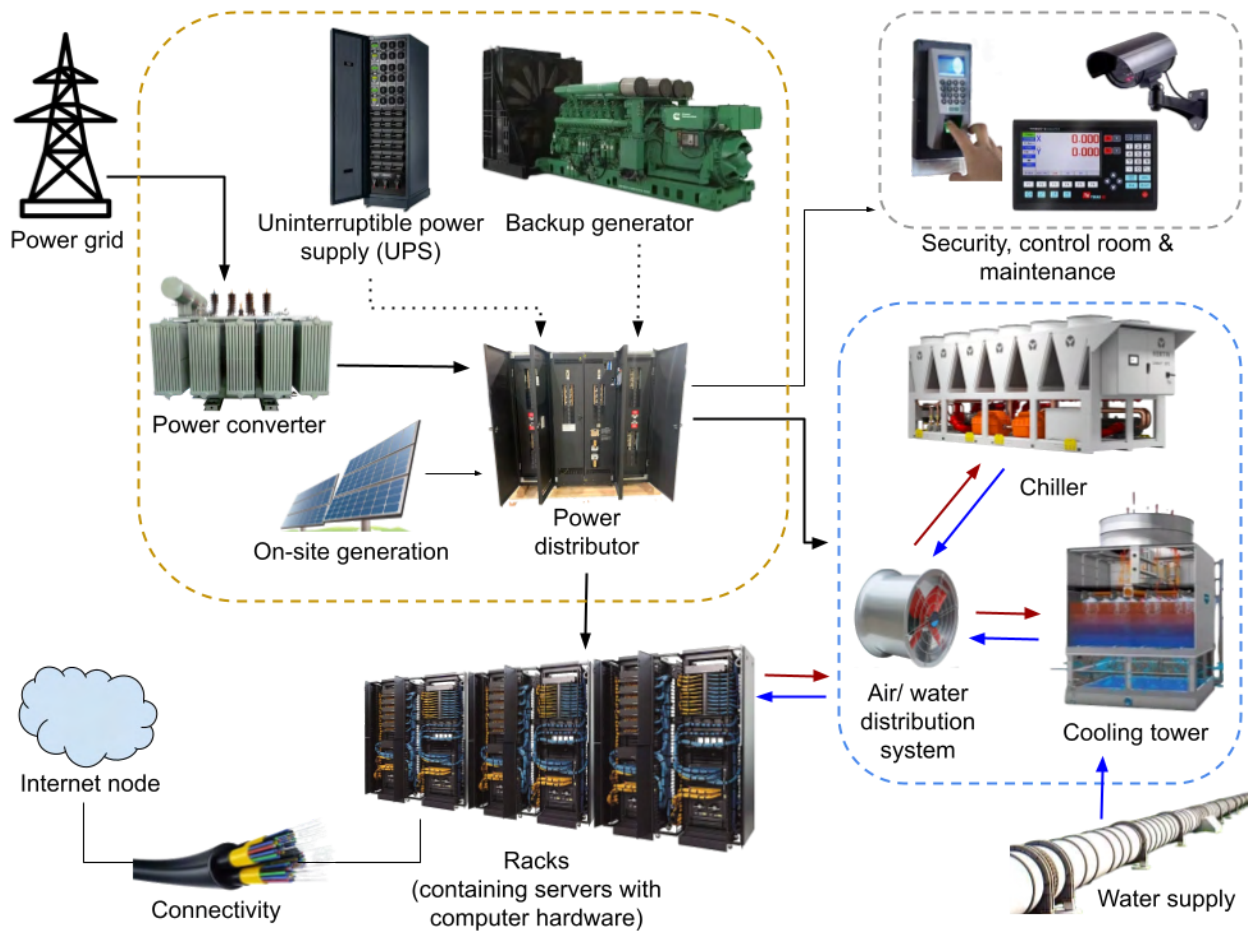

*Figure 2 - Simplified overview of infrastructure components in a data center. Inside the **yellow dashed line**, we see the electrical infrastructure components. The **black arrows** represent the flow of power. A data center receives electricity from the local **power grid** or through **on-site energy generation**, such as solar panels. The incoming power usually needs to be converted from high voltage by a **power converter**. Large data centers have entire substations for this purpose. Electricity is then delivered to a **power distributor**, the central unit regulating power supply to all data center components. In case the power grid fails, data centers typically have batteries installed that can temporarily provide power, called **Uninterruptible Power Supply (UPS)**. If electricity fails for longer than a couple of minutes, **backup generators**, running on diesel, are activated. About 70-90% of the electricity from the power distributor is needed by the hardware itself, contained in **racks** (See Figure 1 for the main components in a rack.). A small fraction of the power is used for **security, the control room, and maintenance (gray dashed line)**, and about 10-30% is fed into the cooling system, displayed inside the **blue dashed line**. **Red and blue arrows** represent the flow of the hot and cold cooling medium, which is typically air or water. The cooling system feeds cold medium into the server room, where it takes heat from the racks. The hot medium is then taken out of the server room by an **air or water distribution system** and fed into cooling components. These can be **chillers** that exchange heat with the outside air or **cooling towers** that cool the medium by evaporating water. As the latter constantly consumes water and is thus connected to a **water supply**.[11] In addition to electricity and cooling, the servers need a fast network connection. Data centers are typically connected directly to **internet nodes** via high-speed **connectivity** fibers. (More in section Key characteristics.)*

---

[11] Cooling systems for data centers differ significantly and are more complex than described here.

# Table of contents

# 1) Background

## Scope

This is a broad investigation into data centers, aiming to provide an overview of how the industry operates, its key actors, and current trends.

As AI is just one of many uses, any given data center may host both AI applications and other services. Therefore, **separating the data center industry into AI/ non-AI parts is infeasible, and this report is concerned with the entire space**.

While other research in compute governance is mainly concerned with the hardware inputs into data centers, **this investigation focuses on the data center infrastructure level**, i.e., how the systems enabling the efficient use of advanced chips are constructed and operated, rather than how the compute clusters hosted by the data centers work. (See [Key data center terms](Key data center terms) for further disambiguation.)

The term "data center" is used to refer to hardware clusters of various sizes. The report primarily focuses on large data centers with a power capacity of >10 MW.

## Context on the project

In the course of this project, I dedicated about 400 hours to studying data centers, of which I spent 200 writing this report specifically. During the project, I spoke to a number of industry experts and attended an industry conference in March 2023.

My strategy was to understand the industry broadly, exploring various topics without dwelling too long on any. I drew key insights from an unfinished 200-hour project by consultant William Hodkins (cited as "draft report").

The investigation is based on public data that primarily covers smaller companies and the colocation sector. It focuses on the US and Europe; insights likely do not fully apply to the Chinese market.

## Data center's relevance for AI governance

The initial motivation of this project, adopting a notion of AI governance outlined by [Dafoe, 2020](Dafoe, 2020), was as follows:

1. **Computing infrastructure (compute) is a necessary input into AI systems.** More powerful hardware has arguably been the main driver of recent AI advancement and may well continue to be in the future (*[Heim, 2021](Heim, 2021)*).
2. Data centers host large amounts of hardware resources and enable their efficient usage. Given the enormous compute requirements of modern ML systems (hundreds to thousands of high-bandwidth interconnected AI accelerators), **large models are almost exclusively trained and executed from this designated infrastructure.**
3. This makes data centers a central part of the (AI) compute supply chain; they provide compute to the end customer, who then uses it to train and deploy AI models (or further rent it out via cloud services).

4. Data centers can further provide **measures of compute**, i.e., ways to estimate & monitor the capabilities of different actors.
5. Because of their central role in the training of large models—the process most concerning from an AI safety perspective—**their governance could contribute to ensuring the responsible and safety-conscious use of AI.**
6. As the data center industry's growth is required for wider adoption of AI technology, forecasting it informs how quickly the AI market can grow in transformative scenarios.

**Note that this report is not exclusively focused on AI training and inference** and only tangentially covers the semiconductor supply chain (see Scope). For more information on the latter, please refer to this reading list on compute governance.)

*For more thorough coverage of how data centers relate to AI governance, see "An assessment of data center infrastructure's role in AI governance", which builds on this report.*

## Key data center terms

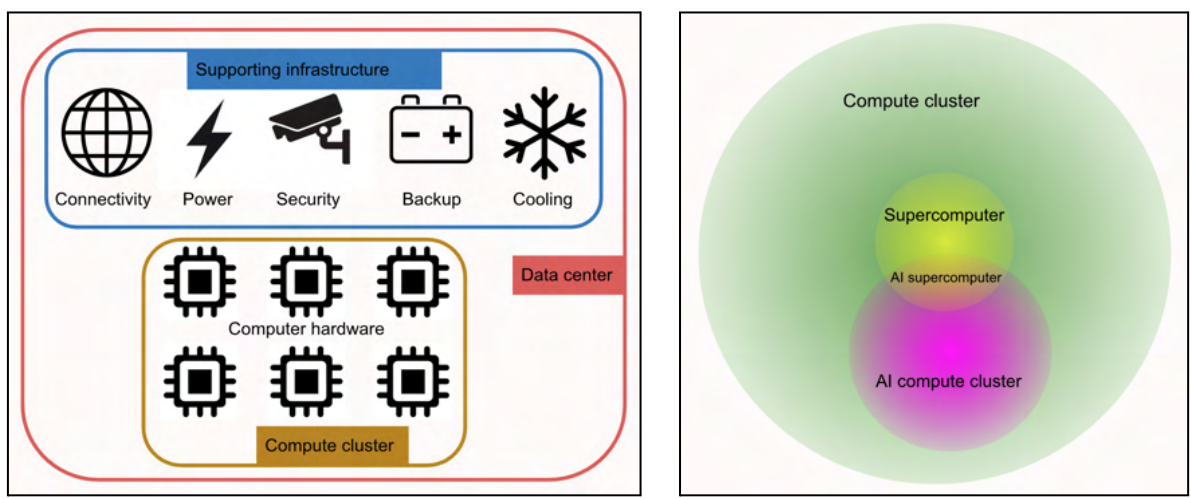

*Figure 3 (left): Visualization of key data center terms. The data center consists of both the supporting infrastructure and the compute cluster, consisting of computer hardware.*

*Figure 4 (right): Venn diagram of compute cluster types. Note that there is no standardized definition of the terms, and the boundaries are continuous. The groups are not to scale (e.g., less than 1% of all compute clusters are supercomputers).*

> There are limited standards and definitions for key terms in the data center industry, and many are used ambiguously for marketing purposes. Based on my review of commonly used phrases (appendix), I tentatively suggest using the highlighted ones. I am open to discussion and think it is reasonably likely that I will change my mind on one or more terms.

A **"data center"** provides the infrastructure for and hosts **"computer hardware"** (short, "hardware") at scale. The data center encompasses both the hardware setup and the infrastructure required to host it.[12]

---

[12] [A data center is] a structure, or group of structures, dedicated to the centralized accommodation, interconnection and operation of information technology and network telecommunications equipment providing data storage, processing and transport services together with all the facilities and infrastructures for power

Most data centers are comparatively small. To specify what type of data center one is referring to, I suggest adding the peak power capacity.[13]

"**Supporting infrastructure**"[14] pertains to the systems required to maintain the hardware, such as cooling, power supply, network infrastructure, backup, and security, but does not include the hardware itself (see Figure 3).

To refer to an aggregation of high-bandwidth interconnected hardware in a data center, the term **"compute cluster"** (also "computing cluster") is used. Note that a data center may host several compute clusters.

A large compute cluster designed to handle demanding workloads and optimized to function as a unified system is known as a "**supercomputer**" or **"high-performance computing (HPC) cluster"**. Both terms are used ambiguously and, depending on the context, may have slightly different meanings. **I suggest using the more common**[15] **term "supercomputer"** and specifying details such as the connectivity bandwidth and system performance to avoid confusion.

Compute clusters typically used for AI training and inference employ high numbers of AI accelerators (primarily GPUs or TPUs) and feature particularly high-bandwidth connections. They are often called a "GPU cluster", though I suggest using the broader term **"AI compute cluster"** to include all types of AI hardware. I further suggest using **"AI supercomputer "** to refer to the largest clusters capable of training advanced systems like GPT-4, or PaLM (see Figure 4). It is important to note that distinguishing AI compute clusters from general-purpose clusters can be challenging without knowing which hardware is employed, as they are housed in similar data centers and set up in comparable servers.

## Units used in the report

**MegaWatt (MW):** The power capacity in MW[16] is one of the most useful proxies for data center compute capacity as specifications for the hardware setup performance (e.g., in FLOP/s) are hardly ever publicly reported.

Examples of power consumption in MW, partly according to Wikipedia:
- 1 MW - 1,000 US households
- 10 MW - typical newly constructed data center; major electric train
- 100 MW - large data center
- 1,000 MW (1 GW) - a metropolis with 1 million inhabitants; the output of a large nuclear reactor
- 19,000,000 MW (19 TW) - world electricity consumption

However, using MW as a proxy for the hardware installed has several limitations:

---

distribution and environmental control together with the necessary levels of resilience and security required to provide the desired service availability (ISO/IEC 22237-1:2021)
[13] E.g., 10MW for a large data center.
[14] Sometimes also "data center facility" or just "facility", though inconsistently used.
[15] See appendix.
[16] Note: Many sources use the total amount of work (energy converted in a given timespan), e.g., kWh. This report instead uses momentarily converted energy in MW. To convert the values, one simply needs to multiply them by the hours the power was used (read this for disambiguation).

1. Computational performance is getting more efficient over time, i.e., for the same amount of energy, one achieves a higher performance for each new generation of hardware.
2. Different data centers use hardware of different efficiency levels[17], so this is a low-resolution proxy.
3. It is important to disentangle the power capacity of the hardware from the power consumption of the data center. The latter may also include power used for cooling and other infrastructure, as well as spare power capacity in case the primary power supply fails. Further, the hardware usually does not run at full utilization, thus typically consuming less power than it would if running at capacity.

**Square meter (sqm)**: Another useful proxy for a data center's compute capacity is its size in square meters. However, the reported space can be equipped with hardware to different degrees, and utilization of the hardware differs as well. When the size is publicly reported, it is often ambiguous whether it is a) the total data center campus b) the building space or c) exclusively the space where hardware can be installed.

For illustration: A typical warehouse is about 2,500 sqm, a football pitch is about 7,000 sqm, and a large aluminum factory reaches up to 80,000 sqm. The largest factory complex in the world, the VW factory in Wolfsburg, Germany, reaches 65,000,000 sqm.

(Back to summary)

---

[17] As hardware is getting more efficient over time.

# 2) What are data centers?

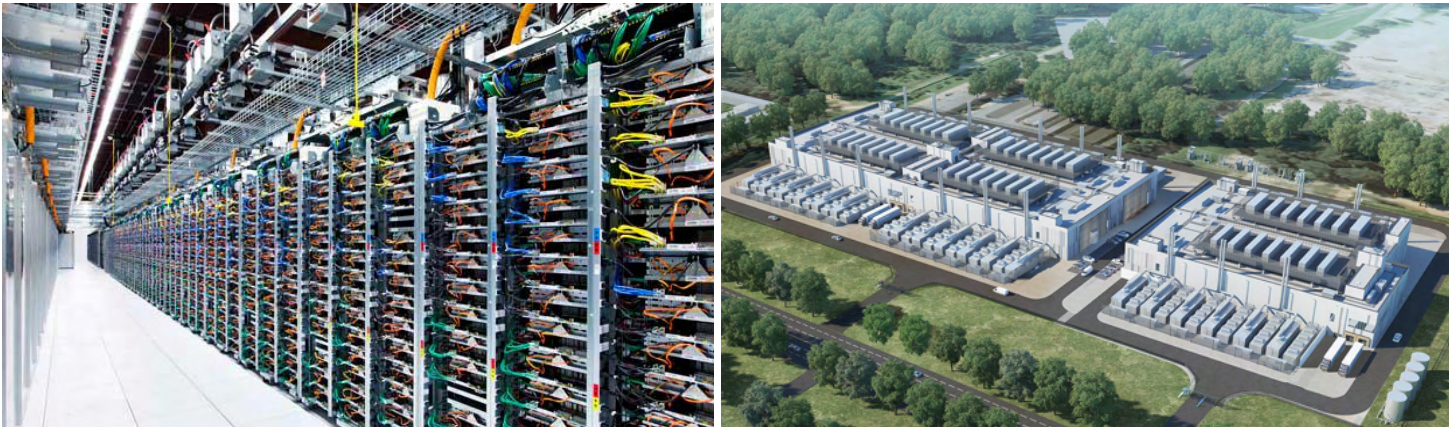

*Figure 5 - A data center from the inside and a large campus from the outside.[18]*

Data centers are purpose-built facilities designed to efficiently host hardware on a large scale in a physically secure environment. A typical data center contains hundreds of server racks, which contain the hardware powering the services it provides. (See Figures 1 and 2 in the summary for a visual explanation.)

AI training is just one application among many. Data centers primarily provide internet services such as hosting websites, online banking, communications, gaming, cloud computing, and data storage.

Physically, data centers are usually large industrial buildings of one or two stories with heavy computer hardware inside, accompanied by large-scale cooling and power infrastructure. They are of comparable size to warehouses, spanning up to several football pitches.

There is no clear definition of a data center; the term is used for single rooms in an office building hosting a dozen servers and the enormous data centers of tech giants alike.[19] The most significant data center projects cost more than a billion dollars and use an area of several dozen football pitches (*Data Center Dynamics, 2022*; *Meta, 2023*).

This report primarily focuses on the **supporting infrastructure** that is provided by the data center to efficiently host the hardware. However, in this report, the term "data center" refers to both the compute cluster as well as the supporting infrastructure (see Key data center terms).

## Key characteristics

Outlines the most important aspects of data centers to provide a basic understanding of how they operate. Note that the report repeatedly refers to these aspects in later sections on Key Inputs for data center construction and operation and Potential limiting factors.

---

[18] Sources: RCRWireless, Vantage.
[19] Unless specified otherwise, this report is concerned with bigger and more modern data centers with a power of >10 MW.

### Power consumption

Modern data centers typically consume tens of MW of power, a key cost driver (see Operating costs). The largest campuses exceed 100 MW, equating to a medium-sized city's use (*Energy Innovation, 2020*). While hardware consumes the most power, 10-40%[20] goes to supporting infrastructure, especially power conversion and cooling (*Statista, 2022a*).

### Cooling

Modern hardware produces large amounts of heat—typically hundreds of watts for a current AI accelerator,[21] accumulating to kilowatts or even megawatts for large compute clusters (*NVIDIA, 2023*). Hot temperatures increase the degradation of installed hardware and risk fires. Therefore, most data centers are operated at just above room temperature, between 18 and 27°C (64 - 81°F). This requires large, expensive cooling systems that continuously absorb the heat produced (*TechTarget, 2022*). Such cooling systems can run on air, water, or special dielectric fluids, and there is a wide variety of systems. For an explanation on different cooling systems, see the section on Cooling equipment.

Cooling consumes significant amounts of power, up to more than ten percent of the total data center power consumption (*Statista, 2022a*). Cooling systems further evaporate large amounts of water (see section on water).

### Uptime and redundancy

Providing internet services and performing high-performance computing both require high reliability and, thus, uptime (the proportion of time a service is available). For large-scale cloud providers, even minor interruptions leading to downtimes can result in substantial financial losses, ranging from hundreds of thousands to millions of dollars (*Data Center Knowledge, 2023*).

To prevent downtime, data centers employ a variety of redundancy components (*CoreSite, 2021*). These include:

1. Backup systems for cooling, power distribution, and network infrastructure to support the hardware in case the primary systems fail.
2. Emergency power supply, including batteries and diesel generators, to power the data center in case of a grid failure.
3. The largest data centers even feature water reservoirs in case the local water supply fails.[22]

In the standard typology for redundancy, 2N indicates a data center has a complete mirror of all power, networking, and cooling systems so that it can afford a full failure of the primary system.[23] For a classification of different levels of uptime, see How can we categorize data centers?.

### Security

Due to the expensive hardware they host, customers' expectations of high uptime, and the high costs of data losses, data center operators invest significantly in physical security.

Typical security measures include (*ISA, 2020*):
- Protection against unauthorized access.

---

[20] Smaller data centers typically need proportionally more power for their supporting infrastructure. Meanwhile, the infrastructure of state-of-the-art data centers can consume as little as 5% of the total power (Google, 2023).
[21] E.g., NVIDIA's H100 has a thermal design power of 700 W (NVIDIA, 2023)
[22] Based on observations of satellite images of large data center projects.
[23] See this article by CoreSite for a thorough explanation.

- Barbed-wire fences and vehicle controls at the campus entrance.
      - Controlled access at the entrance and inside buildings through biometric scanners & documentation of personnel movement.
      - Camera surveillance outside and inside the data center.
      - 24/7 security staff.
      - Remote location.
      - Regular red teaming exercises.
- Protection against other threats.
    - Fire protection systems.
    - Location choice: Construction in areas not affected by floods, earthquakes, or storms.

### Connectivity

Connectivity is the ability to provide fast connections a) between different components within the data center and b) between the data center and the customer. Two key aspects of connectivity are latency and bandwidth.
- **Latency** refers to how long it takes for data to get from the sender to the target. Many internet services require low latency, such as real-time trading, online gaming, and lifestreams.
- **Bandwidth** refers to how much data can be sent per unit time. As an analogy, latency is the amount of time the post service requires to deliver a letter, and bandwidth is the number of words you can write in a letter.

a) Within the data center, network infrastructures are optimized so that servers can communicate efficiently, a process that requires special skills and expensive equipment.

b) Between the data center and the customers, low latencies are achieved by being connected to internet nodes, parts of the global network that have high-speed connections to various locations.

Although most AI training happens offline, such as unsupervised and supervised learning as well as reinforcement learning in simulated environments, some online learning techniques may require internet access. If training occurs across several data centers or required data is unavailable locally, data centers require high bandwidth connections. Furthermore, the same compute clusters used for training are typically also used for inference, which requires a low-latency connection to the end user to provide services in real time, such as voice assistants, image classification, or text generation.

### Complex supply chain

Due to the various systems required in a data center, construction requires a range of specialized inputs, often in high demand and with long waiting times. Although few components are highly specialized, the sheer number of different inputs makes data center projects difficult to manage. (For more, see [Key Inputs for data center construction](#)).

## Types of data centers

This section introduces some common ways of defining different industry segments.

> There are limited standards and definitions for key terms in the data center industry, and many are used ambiguously for marketing purposes.

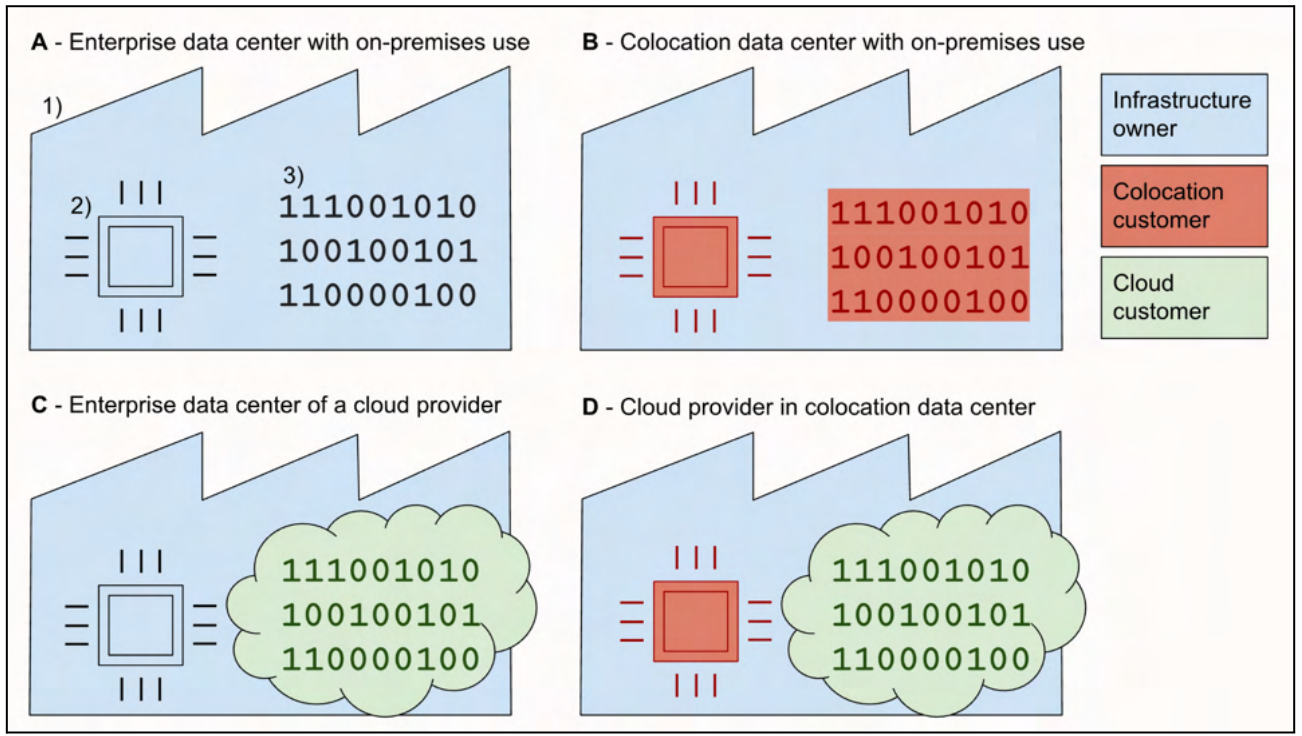

*Figure 6 - Different ownership models of data centers*
*1 - Supporting infrastructure (including the building and various systems for power, cooling, connectivity, security and backup); 2 - Hardware; 3 - Digital resources A - **Enterprise** data center: The Infrastructure owner hosts their own hardware and uses the digital resources themselves. B - Classic **colocation** data center: The infrastructure owner (colocation provider) hosts customers' hardware. The customers use the digital resources themselves. C - **Cloud provider** in an **enterprise** data center: The infrastructure owner hosts their own hardware in an enterprise data center and rents out the digital resources online to cloud customers. D - **Cloud provider** in a **colocation** data center: The colocation provider hosts the hardware of the cloud provider, who rents out the digital resources to end customers.*

Data center ownership: Company-owned, "enterprise" vs. rented, "colocation"

*See Figure 6 for visualization.*

**Enterprise data centers** represent facilities that are owned and operated by individual organizations, such as companies or governments, for the purpose of hosting their own hardware. Although managing these data centers demands substantial effort in terms of planning and coordination, this model can be cost-effective for organizations that have significant and predictable data traffic. There may also be scenarios where owning an enterprise data center becomes necessary due to specialized infrastructure requirements or the need to process sensitive data.

Uncertain estimates suggest that 50-70% of all hardware is hosted by enterprise data centers (*[451 Research, 2019](#)*). The majority of these data centers are smaller in size, typically about a tenth of the size of a colocation data center. Meanwhile, a handful of large cloud companies account for most of the capacity in the enterprise segment. These companies own data centers that are considerably larger than the average enterprise data center, and their share of total hardware is currently on the rise (see [Relevant actors](#)).

The construction and outfitting of enterprise data centers do not necessarily have to be commissioned by the owning company. Some construction firms offer fully operational data centers for sale, a practice known as providing a "**turnkey**" data center.

On the other hand, **colocation data centers**, also known as "multi-tenant", are the go-to option for smaller institutions for whom constructing their own facilities is not cost-effective. Instead, these entities lease capacity from colocation providers, who manage the entire supporting infrastructure. This includes power supply, cooling, connectivity, security, backup systems, and to some extent, hardware maintenance. Customers of these services simply install their hardware in fully prepared racks (*[TechTarget, 2020](#)*).

Globally, colocation data centers are estimated to host about 30-50% of all hardware. These centers typically provide rack space in a shared room, a service known as "**retail**," or even entire floors or buildings, referred to as "**wholesale**." Most colocation providers also offer pre-installed hardware for purchase by the customer, in addition to the option for customers to install their own hardware.

My current impression is that large AI compute clusters are predominantly hosted in enterprise data centers, and large cloud providers typically almost exclusively rely on their own data center fleet. However, an industry expert suggested that at least Microsoft Azure frequently uses colocation data centers when expanding their services to new regions. Furthermore, NVIDIA offers certifications for [AI-ready data centers](#), indicating that at least some AI compute clusters are hosted in colocation facilities.

## Who uses the digital resources?

Another meaningful way to separate the data center industry is per the end user of the digital resources the data center produces.

**On-premises** usage indicates that digital resources are used in-house, i.e., by the company owning the hardware. Note that this is independent of the data center ownership model, so on-premises compute can be hosted both in an enterprise as well as a colocation data center.

**Cloud** (off-premises) usage indicates that digital resources are rented out online. The end user owns neither the data center nor the hardware (and usually does not know where either is located) (*[Pure Storage, 2022](#)*).

Within the cloud segment, **hyperscalers** build and own, particularly large data centers. While the term is not well defined, it usually refers to major cloud companies and their data centers (> 10 MW), such as AWS, Google Cloud, Microsoft Azure, IBM Cloud, and Alibaba Cloud. Due to their high capacity, these companies offer highly scalable resources and are attractive to startups anticipating quickly growing demand.

There are also **hybrid** forms, where parts of the compute are used by the owner, and the excess is rented out online.

Cloud services can further be distinguished into three categories (Figure 7) (*[IBM, 2022](#)*):
1. Infrastructure as a Service (**IaaS**), providing **virtual machines**, which are independent operating systems run on the cloud infrastructure. While this leaves a high degree of flexibility to the customer, it also requires significant setup and maintenance efforts. Example providers include AWS and Google Cloud.
2. Platform as a Service (**PaaS**), providing a **complete environment** where the customer can install their own software. It leaves less room for customization than IaaS but is also much easier to manage. Example providers include Microsoft Azure and Google App Engine.
3. Software as a Service (**SaaS**), providing **applications** that can be directly used by the customer with very few setup requirements. E.g., online mail services and management tools. These are usually not referred to as cloud computing but rather called **internet services**.

Within the context of this report, whenever "cloud" is used, it primarily implies IaaS.

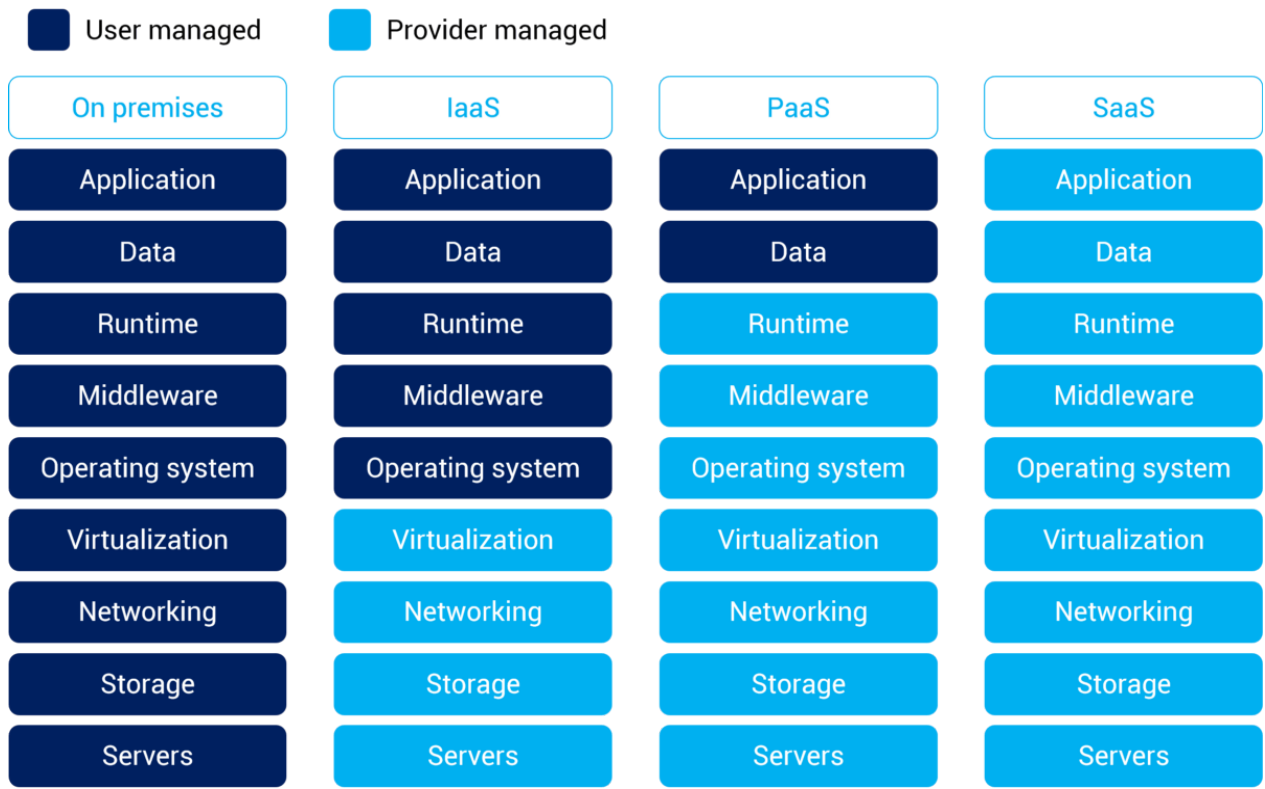

*Figure 7 - Models of cloud computing ([Alibaba Cloud, 2022](#))*

Other categorizations

These are some additional classifications of data centers not used in this report.

Depending on the **level of redundancy**, data centers have different guaranteed uptimes (percentage of time the service offered by the data center is online); these are categorized into **different Tiers,** with Tier 1 being the lowest (99.671% → ~one-day downtime per year) and Tier 4 being the highest (99.995% → ~half an hour of downtime per year) (More on [Uptime Institute](#), an industry auditing company certifying these Tiers.).

**Latency requirements:** **Edge** (sometimes "micro") data centers are optimized for especially low latencies, which are needed for, e.g., high-speed trading or controlling autonomous vehicles. They are comparatively small, expensive, and located close to the end user (*[TechTarget, 2022](#)*).

For other data centers, latency is similarly an important factor, particularly for cloud computing, streaming, or communication. Many are built close to metropoles despite the higher land and energy costs. Some use cases, such as data archiving, have relaxed latency requirements, and the corresponding facilities can be built in the countryside.

**Suitability for AI training:** Currently, there is no taxonomy for data centers used mainly for AI applications. Industry experts report that because AI accelerators require more electricity and cooling power than general-purpose compute (CPUs), AI applications are usually hosted in larger data centers. However, single-purpose data centers exclusively used for AI training are uncommon. Since the industry currently does not differentiate between AI and non-AI data centers, this report focuses on the broader data center space unless indicated otherwise.

**Use cases**: Data centers host a wide variety of services, including:
1. Data storage and archiving, e.g., Dropbox, GitHub[24]
2. Public services, e.g., medical information, tax software
3. Communication services, e.g., e-mail, video calls, messaging
4. Trading and e-commerce, e.g., Amazon, online banking, stock exchange
5. Productivity tools, e.g., online editors, management software
6. Online gaming
7. Streaming services, e.g., YouTube, Spotify
8. High-performance computing (HPC)
    - Scientific simulations
    - Large-scale data analysis
    - Machine learning training and inference, e.g., voice assistants, ChatGPT

## How many data centers are there?

It is surprisingly difficult to estimate the total number of data centers of different sizes. While colocation data centers are reported systematically, information in the enterprise segment is rare.

Based on adding the number of data centers reported by big tech companies and a confirming estimate via the global power consumption of data centers, there are **an estimated 140 large data centers** with a total power capacity of >100 MW globally (70% confidence interval **110 - 225**). See How many data centers have a capacity of > 100 MW? in the appendix for the estimation approach.

Based on the estimate above, there are likely **~400 big data centers with a power capacity of 10 - 100 MW** (70% confidence interval **225 - 1,100**).[25] This is the data center size most likely to host current AI compute clusters.[26]

Including small data centers, there are 10,000 - 30,000 (70% confidence interval) data centers with a maximum capacity of >0.1 MW.
For reference:

---

[24] It is not useful to divide the space into data storage and data processing as most data centers offer both (Although there is a niche for data archives, where the data is hardly ever accessed)
[25] An industry expert confirmed that this is likely the right order of magnitude.
[26] E.g., the Leonardo supercomputer in Italy, containing around 13,000 NVIDIA A100, has a power capacity of about 10 MW. Industry AI compute clusters most likely occupy only parts of larger data centers.

- [Statista, 2023a](#) reports just below 5,000. However, it is unclear if this includes enterprise data centers, and they do not report their sources.
- [Datacenters.com](#) reports about 2,200 colocation facilities as of March 2023 Assuming that colocation accounts for 40% of all data centers, this implies about 5,500 data centers globally.
- [451 Research](#) reports more than 8,500 for the colocation and wholesale segment as of March 2023. If the segment includes 40% of all data centers, the total is 21,250.

I expect a significant number of data centers will not be covered in these estimates, especially smaller enterprise data centers and data centers in China.

For the geographical distribution of data centers, see [Locations of data centers](#).

([Back to summary](#))

# 3) State of the industry

Describes the main features of the industry, such as market growth, consolidation, and distribution of data centers.

## Market size and growth

*The figures presented rely entirely on public data. Industry experts suggested that market data is often unreliable.*

The global data center market is valued between **$206B and $321B** (*GlobeNewswire, 2022*; *Statista, 2022b*)[27]. These figures comprise the facilities, their infrastructure, and services directly associated with them. For reference, the global semiconductor market is two to three times larger, at ~$600B (*Fortune Business Insights, 2022*).

The market has grown by about 26% in the last five years (*Statista, 2022b*)[28]. Analyses expect the market to grow at a **CAGR of 10%**, meaning it will double in the next seven years (*GlobeNewswire, 2022*; *Kanhaiya et al., 2021*, *McKinsey, 2023*); Technaivo, 2022 even estimates a CAGR of >20%, though Statista, 2022b only estimates around 5% of growth per year. Surprisingly, despite being a crucial input, the server market's growth is projected with a lower CAGR of 5 - 8% (*imarc, 2022*; *Grand View Research, 2020*; *Statista, 2022c*).[29]

**AI Infrastructure** is one of the fastest growing markets, currently estimated at ~$30B and expected to rise above $100B in the next five years. (*MarketsandMarkets Research, 2022b*; *KBV Research, 2022*; *Mordor Intelligence, 2022*). Note that definitions for this market vary. Meanwhile, the **cloud market**, which focuses on the services provided rather than the infrastructure (but is not always cleanly separated from the data center market), is valued at around $450B (*MarketsandMarkets Research, 2022b*; *Statista, 2023d*)

For a regional breakdown of the market, see Per country. For potential future bottlenecks, see 5) Potential limiting factors for industry growth.

## Relevant actors

### Most important companies

Globally, the **colocation** market is divided across many actors, most of which have less than 1% of the global market share (Figure 8) (*Statista, 2022d*). The three biggest actors' share has risen by six percentage points since 2016 (*451 Research, 2017*). The market will likely continue to get more concentrated as large actors improve their economies of scale.

---

[27] This number excludes chip production.
[28] The numbers may not be adjusted for inflation.
[29] There seems to be no clear explanation for this disparity other than that market definitions or methodologies may be different between this estimate and estimates for the whole industry.

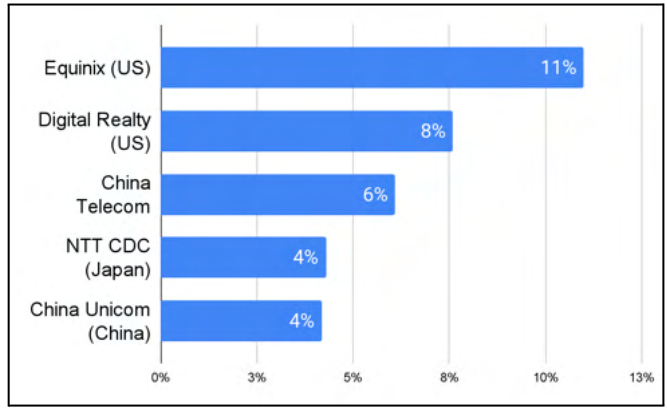

*Figure 8 - Major companies of the global Colocation Market as reported by Statista, 2022d.*

The global **cloud market** is dominated by Amazon Web Services, Microsoft Azure, and Google Cloud (Figure 9). Other important companies include Alibaba (5%), IBM (3%), Salesforce (3%), Tencent (2%), and Oracle (2%). The market is getting more concentrated over time, with the top actors already accounting for about two-thirds of the volume (Statista, 2023).

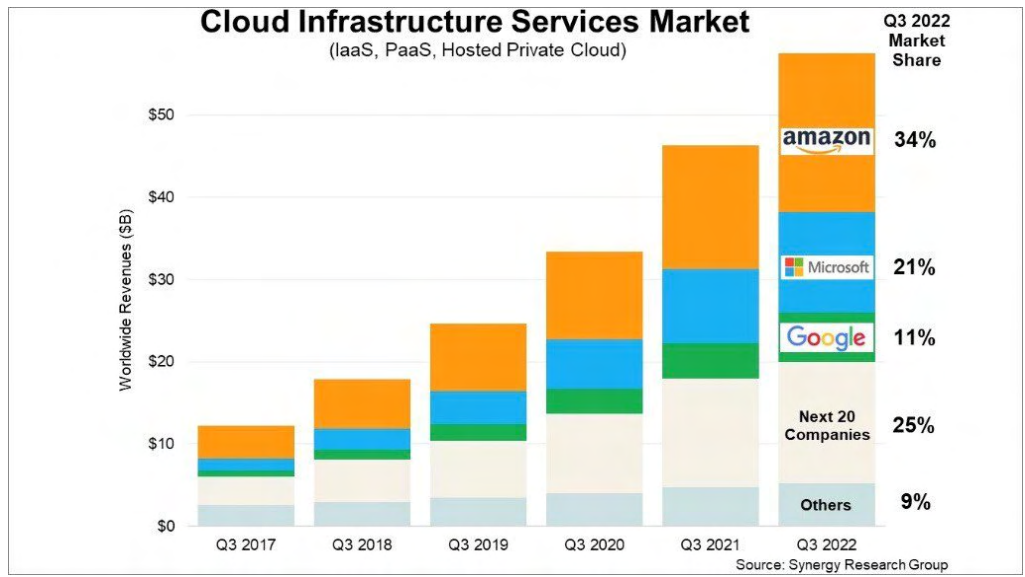

*Figure 9 - The cloud computing market, according to Weissenberger, 2022.*

Meanwhile, the **cloud AI market** is valued at $5B[30] (Mordor Intelligence, 2022), with top players including AWS, Microsoft, Google, IBM, and Intel. Allied Market Research, 2023 further lists Apple, Baidu, Cisco, Nuance Communications, SAP, Tencent, and Wipro. An industry expert claimed that the cloud AI market is growing at an unsustainable pace of 80% per year. However, no public sources confirm this.

---

[30] Intuitively, this seems like an underestimate. However, few public estimates exist.

Government actors

A variety of government entities, particularly within the United States, utilize a broad range of data centers of differing sizes. Reports from [The Register](#) suggest that there are several thousand such data centers, although the majority are likely to be smaller builds.[31]

One notably recognized government facility is "[the Bumblehive](#)", a high-security, high-redundancy data center campus that costs around $1.5B and has a power capacity of 65 MW. Although other governments may similarly own custom data centers, government data centers likely represent a minor proportion of the global capacity.[32]

Not all government workloads are hosted on-premises. For instance, AWS provides a service called [AWS GovCloud](#), which has been specifically designed to satisfy the security and privacy standards of governments and their contractors. A notable aspect of this service is that it is exclusively managed by US citizens.

## Locations of data centers

I describe where data centers are typically built and which countries have the highest number of them.

### Per country

With about a third of all data centers, the United States is currently the most important data center location by a considerable margin. However, the exact proportion remains uncertain due to the considerable variance found in different estimates, many of which do not specify the sizes of the included data centers (*[US International Trade Commission, 2021](#)*[33]).

It is particularly difficult to estimate China's capacity. While it may account for the second-highest number of data centers, most sources suggest that the United Kingdom and Germany rank higher (*[Statista, 2023a](#)*). However, since considerably less public data exists on Chinese data centers, there is likely a high number of unreported data centers in the PRC.

In terms of revenue, the United States is the most substantial market, trailed by Europe and China. All three markets are growing at similar rates, with a projected increase of $20B in revenue over the next five years.
- US: $95b, 30% of global revenue (*[Statista, 2022e](#)*)
- Europe: $79b, 25% of global revenue (*[Statista, 2022f](#)*)
- China: $64b, 20% of global revenue (*[Statista, 2022g](#)*)
- Other markets with more than $5B in revenue include Japan ($16b), India ($7b) and Korea ($6b). (*[Statista, 2022b](#)*)

Certain countries, such as Russia, China, and Turkey, enforce data sovereignty laws, mandating that their citizens' data be stored domestically[34]. This legal requirement prompts the construction of a higher number of data centers in these countries (*[US International Trade Commission, 2021](#)*). In the case of China, restrictions on data traffic across borders (in terms of both bandwidth and content) further necessitate hosting most services within the country.

---

[31] For more information, see this report by the [US GAO](#).
[32] In principle, there could be significant government or military data center capacity that remains concealed. However, there is little evidence indicating this.
[33] The report cites CloudScene, though does not provide a direct link, so the the source is unclear.
[34] For an overview, see this [Wikipedia](#) list.

Much like many other infrastructure projects, China has a national strategy for data center construction. The strategy promotes the construction of data centers in national hubs, including four mega-projects located in the less populated, power-abundant western parts of the country. Rather than investing directly, it instead seems like the central government seeks to guide the investment decisions of private companies and regional authorities (*[China Briefing, 2022](#)*; *[Reuters, 2021](#)*).

Geographical distribution

Colocation data centers, for which systematic reporting of information is available, are typically constructed in close proximity to metropolitan areas. This strategic positioning enables these centers to provide their customers with low latency. Some significant regions for these types of data centers, according to public data primarily from colocation providers, include ([datacenters.com](#), [baxtel.com](#), [datacentermap.com](#)):
- The DC area (including northern Virginia), which has the largest concentration of data centers worldwide. Factors contributing to this concentration include inexpensive energy, connectivity to cities along the East Coast, and the presence of numerous contractors in the DC area (spanning sectors such as intelligence, defense, and finance) (*[Data Centers Today, 2020](#)*)
- Other significant areas in the United States include Dallas, the Bay Area, Los Angeles, Chicago, and the New York area.
- Internationally, key areas include London, Amsterdam, Frankfurt, Hong Kong, Singapore, and Sydney.

On the other hand, enterprise data centers, especially those owned by major companies like Google, Meta, Microsoft, and AWS, tend to be situated further from metropolitan areas (based on locations reported by [Meta](#), [Google](#), and [datacenters.com](#)). The locations of these data centers are influenced by different factors compared to colocation providers. Enterprise data centers are typically larger than colocation facilities, thereby requiring more space and power. Moreover, due to their economies of scale, these companies can afford to establish new power lanes or connectivity cables (*[Google, 2018](#)*).

In contrast, colocation facilities aim to meet a diverse range of customer needs and therefore have less flexibility in their choice of location. These facilities are incentivized to be within easy reach of their customers, which often results in their proximity to large cities.

Overall, the most important factors for data center locations are:
- Energy prices/ abundance of reliable power supply and increasing availability of renewables as a consequence of sustainability requirements
- Location of internet cables (See [this map](#) for an overview)
- Proximity to important markets
- Frequency of disasters (No data centers in areas with frequent earthquakes, storms, floods, etc.)
- Beneficial cooling conditions such as a cold environment and water abundance.

Location trends

Bigger data centers are increasingly constructed further away from metropoles (*Data Center Frontier, 2022a*; industry professionals). This shift is not only due to these centers' high power consumption but also because building permits are typically more readily obtainable in less economically developed regions. Furthermore, for large-scale projects, establishing new power and connectivity infrastructure becomes a viable option, making remote locations more feasible.

There are reports suggesting that data centers are migrating north to locations such as Canada and Scandinavia (*Computer Weekly, 2021*). However, it seems more likely that this trend represents an expansion rather than a major shift in the industry.

In certain locations, data centers consume so much power that governments restrict further construction. E.g., in Ireland, data centers consume about 18% of the country's electricity, leading to a hold on new projects. (*Euronews, 2023*; *Data Centre Dynamics, 2022*).

The necessity for water somewhat limits data center locations (*Data Center Frontier, 2023*). However, if energy is abundant, data centers can rely on alternative cooling sources that do not require a constant water supply.

In a speculative vein, as AI increasingly becomes a matter of national security[35], countries may begin efforts to reduce their dependence on foreign data centers. The GAIA-X project, initiated by the European Commission in 2019 to bolster digital sovereignty, could be seen as an early indication of this trend. However, the current impact of GAIA-X is unclear, and there is uncertainty about the potential for a significant push toward more regional infrastructure beyond government and military applications. If a push for greater digital sovereignty does occur, it could result in a more globally distributed network of data centers, even though the United States would likely continue to host the highest concentration of facilities.

Shift to the cloud

According to Gartner, 2022 there is an ongoing trend of businesses shifting their applications to the cloud, projected to increase from 41% in 2022 to 51% in 2025. This shift is accompanied by an uptick in businesses' spending on cloud services. Industry experts also cite a growing tendency for companies to rent AI compute resources rather than owning them due to the rapid pace of innovation in this field. The frequent need to update hardware, which can be both costly and labor-intensive, contributes to this trend.

The shift toward the cloud is particularly prevalent in the AI industry. Major AI labs like OpenAI and Anthropic established partnerships with Microsoft and Google, respectively (*Anthropic, 2023*; *Microsoft, 2023*), and AWS recently announced a partnership with Hugging Face (*Hugging Face, 2023*). Meanwhile, Google DeepMind uses Google's internal cloud compute.

This increasing reliance on large cloud providers is due to the increasing intricacies of setting up and maintaining AI compute clusters, requiring significant tacit knowledge that primarily exists at large cloud providers. Further, in a bid to dominate the emerging market, big cloud companies such as AWS are known to offer highly competitive pricing to AI labs, even at the cost of not making a profit in the early years.[36]

---

[35] E.g., because of further revelations such as the Snowden leaks or because there are important military applications of AI.

However, there is a countervailing trend with some companies moving their AI workloads back to on-premises infrastructure. The reasons behind this shift include higher costs associated with cloud usage, latency issues that can slow down AI applications, and data privacy or sovereignty concerns (*Protocol, 2022*; *Data Center Frontier, 2022b*). Despite these concerns, the overall trajectory seems to favor a move towards cloud-based solutions, with the AI cloud market projected to more than double in the next five years (*Mordor Intelligence, 2021*).

([Back to summary](#))

---

[36] According to an industry expert.

# 4) Building & running a data center

This chapter adopts a data center operator's perspective to understand the requirements and limits of the industry.

## Key inputs for data center construction

This section covers the main categories of inputs required to construct a data center. For an analysis of which ones could be likely bottlenecks, see [Potential limiting factors for data center growth](). Also, see this [visual overview]() of the different inputs.

Suitable sites

Data centers demand considerable space, with an average campus, including its surrounding area, occupying about 10,000 square meters. Larger campuses can expand across up to 700,000 square meters, which is comparable to the area of approximately 100 football fields (*[Dgtl Infra, 2022]()*).

For reference: A typical warehouse is about [2,500 sqm](), a football pitch is about [7,000 sqm](), and a large aluminum factory reaches up to [80,000 sqm](). The largest factory complex in the world, the VW factory in Wolfsburg, Germany, reaches [65,000,000 sqm]().

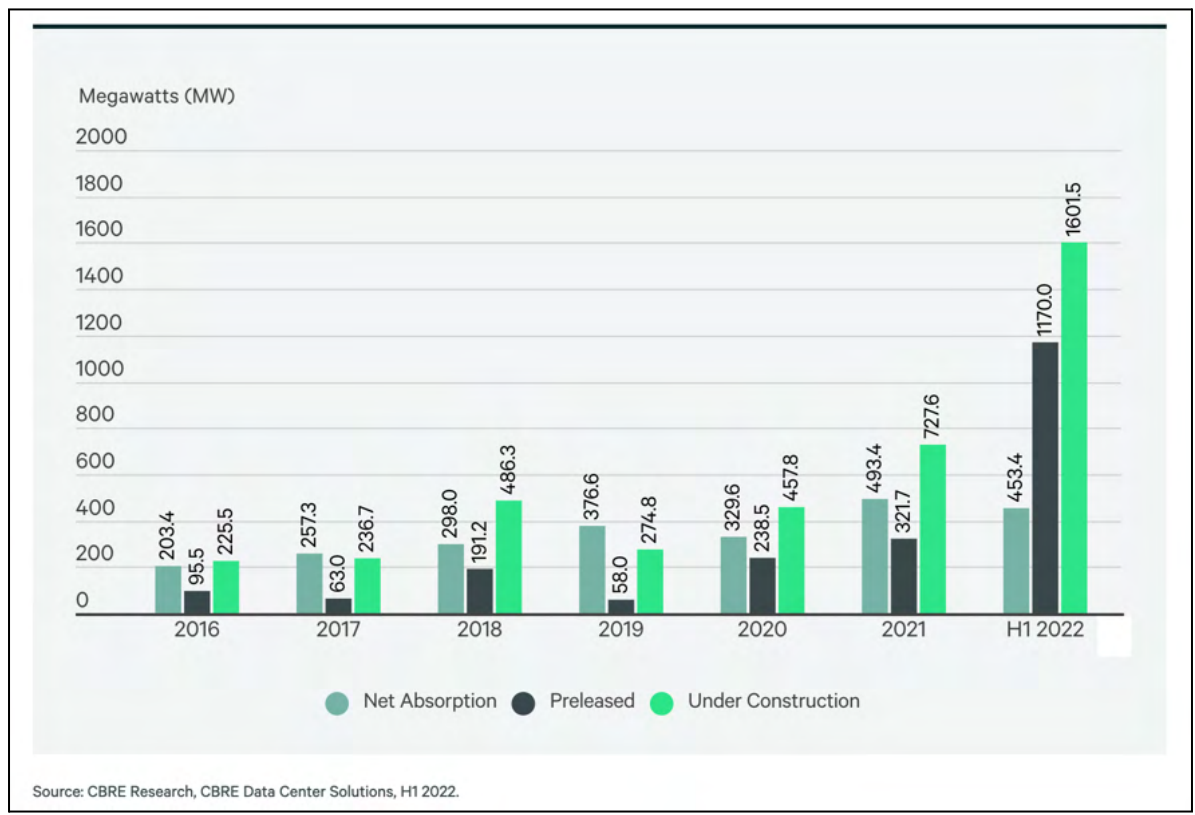

*Figure 10 - power capacity absorption of the North American wholesale data center market according to [CRBE 2022]().[37]*

---

[37] Note that this just represents one market segment.

In addition to the considerable spatial requirements, the availability of suitable sites for data centers is restricted, as they require a number of further prerequisites. These include the availability of spare power capacity in the grid, proximity to internet cables, water for cooling purposes, and a low likelihood of experiencing floods, storms, and earthquakes. (Also see [Geographical distribution](#).) Due to the combination of these factors, finding sites to establish new data centers is currently a considerable challenge ([Data Center Frontier, 2023](#)). A significant contributor to this is a recent increase in power demands of the industry, both because more data centers are constructed and because each data center has a higher individual power consumption (See Figure 10).

Regional legislation also imposes constraints on where data centers can be constructed. For example, the construction of data centers in Dublin was put on hold when their energy consumption levels became excessively high *([Business Post, 2022](#))*.

The costs of land and the taxes levied vary extensively, with these expenses typically being significantly lower in regions situated outside metropolitan areas.

### Construction material

As other industrial buildings, data centers require basic construction materials such as steel, aluminum, and concrete.

### Power infrastructure equipment

Data centers, particularly larger ones, require substantial amounts of power, with consumption levels potentially reaching hundreds of megawatts, akin to the energy demands of large cities. Therefore, the construction of a data center often necessitates the establishment of new power lines to connect the facility to the nearest high-voltage line.

It is not uncommon for data center providers to establish wind or solar parks close to large projects, e.g., [Google added](#) more than 7,000 MW of renewable energy in 2021.

Data centers also necessitate various types of power distribution infrastructure. First, the high voltage needs to be transformed to a usable level, which requires large transformers. Next, surge protectors and the UPS (Uninterruptible Power Supply) ensure a consistent power supply that is immune to fluctuations in the grid. Finally, cooling systems and every server need to be connected to the power source via additional power distribution units. For a comprehensive overview of power infrastructure in a data center, see [Strategic Media Asia, 2020](#).

### Networking equipment

To run real-time applications, data centers necessitate low-latency, high-bandwidth connections. Often, this requires the installation of new fiber-optic lines to establish a link with the nearest high-speed internet junction. Large corporations, such as Google, may even undertake the construction of proprietary, long-distance fiber-optic connections between their data centers. This can encompass the deployment of submarine cables for intercontinental data transmission ([CNET, 2018](#)).

Within the data center, infrastructure like optical fiber cables and network switches are integral to achieving efficient communication between servers. According to an industry expert, for AI compute clusters, the networking equipment to allow high-bandwidth connection can cost almost as much as the hardware itself.

### Backup components

Services provided by data centers, such as online banking, communications, streaming, and computationally intensive tasks like AI training, require high reliability. Consequently, data centers are equipped with backup systems to maintain uninterrupted service under all conditions (*CoreSite, 2021*).

1. Emergency power supply, including **batteries** (as part of the UPS)[38] and **diesel generators**, in case of a grid outage.
2. Backup components for electricity distribution to all necessary components in case of primary source failure. Fully redundant data centers even feature two independent power grid connections.
3. Backup systems for network connectivity.
4. Additional cooling systems in case of failure.

The quantity and complexity of backup components vary, depending on the specific requirements of a data center. The more sensitive the services provided, the more extensive the backup infrastructure needs to be.

### Safety and security equipment

Besides redundancy, data centers ensure reliability by protecting against threats:
- Physical attacks: Cameras, barbed-wire fences, biometric scanners, security doors, and street lights on the entire area
- Fire: Fire detectors and a variety of different fire suppression systems (more on TechTarget, 2022)

### Servers

Data center computations run on servers, which comprise a mix of general-purpose and specialized chips, tailored to the specific applications being provided (See Figure 1 for a visualization). These chips originate from incredibly intricate supply chains, as shown in this visualization: CSET 2021. Although by far the most sophisticated input, semiconductors remain outside the scope of this report.

Besides CPUs and GPUs, servers encompass other critical components such as DRAM (Dynamic Random Access Memory) for storage, network interface cards for efficient communication, and transceivers that convert optical signals into electrical ones. These components, too, are sourced from complex supply chains.

### Cooling system

To prevent overheating and degradation of hardware, it is constantly cooled by cold air, water, or specialized non-conductive fluids.

A typical cooling system consists of two main components, a cooling source, which absorbs the heat from the cooling system, and a distribution system, which moves the heat from the servers to the cooling source (*datacenters.com, 2020*).[39] Some cooling systems further contain additional components, such as dehumidifiers.

---

[38] Uninterruptible Power Supply, see Power infrastructure equipment.
[39] For a further distinction, see datacenters.com, 2020.

Cooling sources

There are different types of cooling sources that can be separated in active and passive heat exchangers (*TechTarget, 2022*).

Passive exchangers dissipate heat simply by bringing the cooling medium into contact with colder outside air (or sometimes nearby bodies of water). This only works as long as the temperature outside is below the server room temperature[40] and is thus mostly used in colder climates.

Active heat exchangers utilize energy or water to cool the cooling medium. For instance, cooling towers leverage the process of water evaporation to remove heat from the cooling medium. While this method only needs small amounts of energy, it consumes significant amounts of water (also see section on water). Meanwhile, chillers function much like household refrigerators, concentrating heat on their surfaces for more efficient heat dissipation. Although this process obviates the need for water, it demands substantial electrical power. A typical data center uses a mixture of passive heat exchange, cooling towers, and chillers, depending on the climate and availability of water.

Active exchangers use energy or water to cool the cooling medium. Cooling towers evaporate water to dissipate heat from the cooling medium—a process that can use large amounts of water (See section on water). Alternatively, similar to a kitchen refrigerator, chillers use refrigeration cycles to concentrate the heat at their surface for faster dissipation. While this process does not require any water input, it uses considerable amounts of electricity.

---

[40] Typically between 18 and 27°C (64 - 81°F), see Cooling.

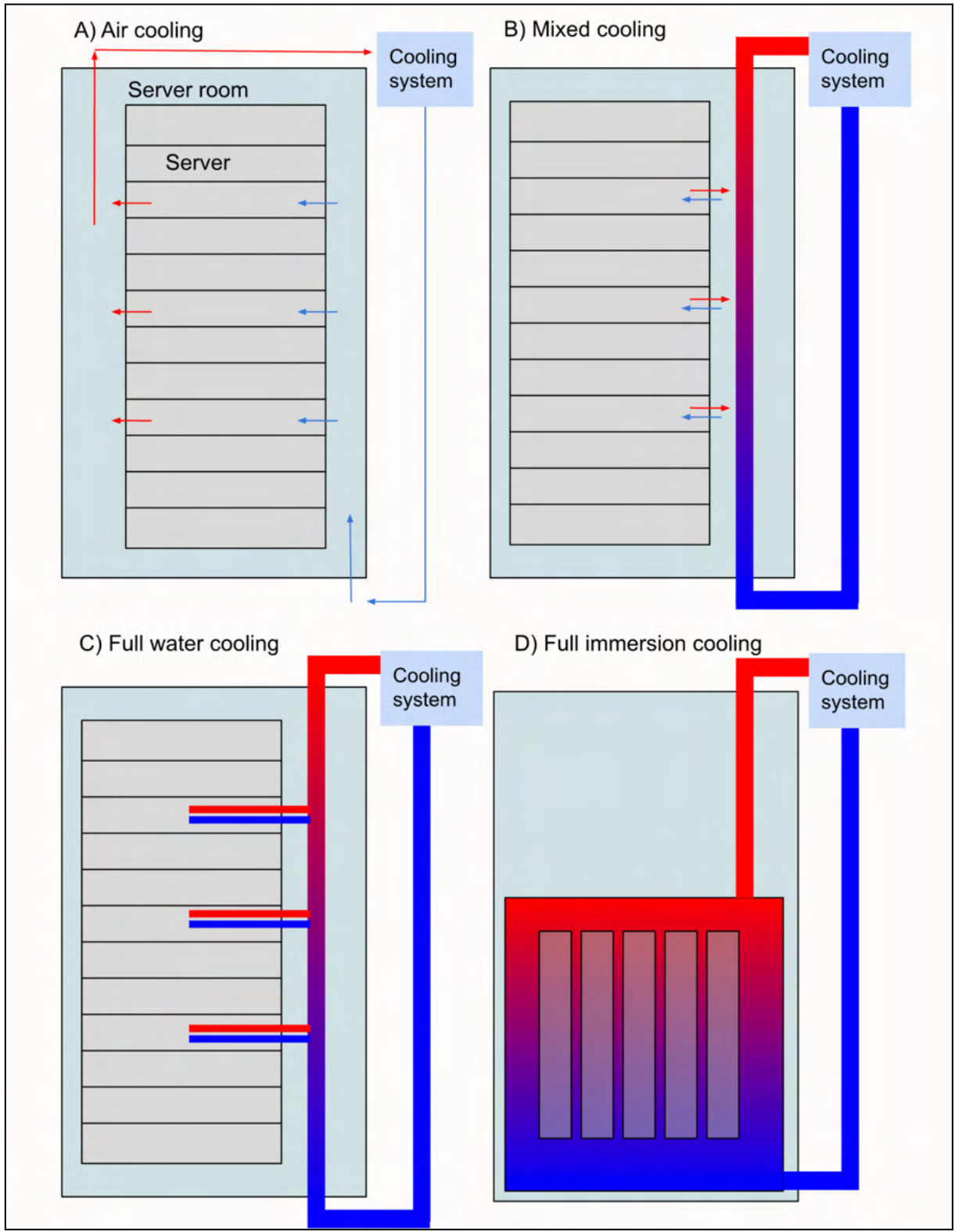

*Figure 11 - Different cooling types inside the server room. Arrows represent air; soli d tubes represent water or other liquid cooling media. The server room is shown in turquoise, server rack in gray. (Thorough Explanation below.)*

Distribution systems

Mostly independent of which cooling sources are used, various solutions exist for moving the heat generated by the hardware there.

Historically, data centers primarily relied on air-based distribution systems for cooling purposes (see Figure 11 A). Recently, however, there has been a shift towards more efficient, albeit complex, water-based systems. One such system is **mixed cooling**, where air absorbs the hardware's heat, which is then exchanged with water pipes situated behind the servers (Figure 11 B). While this method is compatible with a multitude of servers and relatively simple to implement, it does not match the efficiency of full water cooling (*AKCP, 2021*).

**Full water cooling** involves cooling the hardware directly with water. Small tubes carry water straight to the chips, the primary source of heat within servers (Figure 11 C). This system can be challenging to set up and maintain, and to mitigate the risk of equipment damage from potential water leakage, non-conductive fluids are often used instead of water.

Finally, **immersion cooling** involves completely covering the hardware in a non-conductive liquid (Figure 11 D).[41] Upon absorbing a certain amount of heat, the fluid evaporates and rises to the container's top. It can then be directed to an external heat sink. After releasing its heat, the gas condenses and passively flows back into the containers, requiring minimal mechanical energy, making it highly energy-efficient (*Kunkcoro et al., 2019*). However, because servers are stored vertically, different designs are required, making installation and maintenance more complex.

Currently, air-based cooling systems are still dominant, and even high-end hardware, such as NVIDIA's H100, is currently designed to be air-cooled (*NVIDIA, 2023*). However, several consulted industry experts predict the widespread adoption of liquid cooling techniques over the coming years due to the increasing energy density of high-end hardware and stricter requirements for energy efficiency of data centers due to environmental concerns.

## Key inputs for data center operation

### Power

Data centers are significant power consumers, with individual campuses often utilizing anywhere from hundreds of kilowatts to hundreds of megawatts of electricity. Moreover, a growing trend indicates increased power consumption in data centers, with the typical capacity of recent projects being around 20 MW.[42]

In the context of data center energy consumption, a key metric to consider is Power Usage Effectiveness (PUE). PUE is a measure of a data center's energy efficiency, comparing the total power consumption of a data center to the power used solely by its IT equipment. For example, if a data center has a PUE of 1.5, this means for every 1.5 units of power consumed by the facility, only 1 unit is used by the IT equipment, with the rest being used for non-computational tasks such as cooling and power conversion. A lower PUE score indicates a more efficient data center, implying a larger portion of power is used for computational functions rather than supporting infrastructure (*TechTarget, 2022*).

---

[41] For an intuitive understanding, watch this 60-second video.
[42] Quoting the draft report: Recent broker reports claim that there are now "more 20+ MW requirements than 1 MW requirements" and that "more 100+ MW leases will be signed".

Notably, a significant share of a data center's power is lost during power conversion or used by the cooling systems (see Figure 12). However, the application of state-of-the-art cooling systems can significantly enhance data center efficiency. As an example, Google's data centers use as little as 5% of total power for conversion and cooling (*Google, 2023*). This figure represents a broader trend in the industry, where the PUE is generally decreasing over time (*Statista, 2022j*).

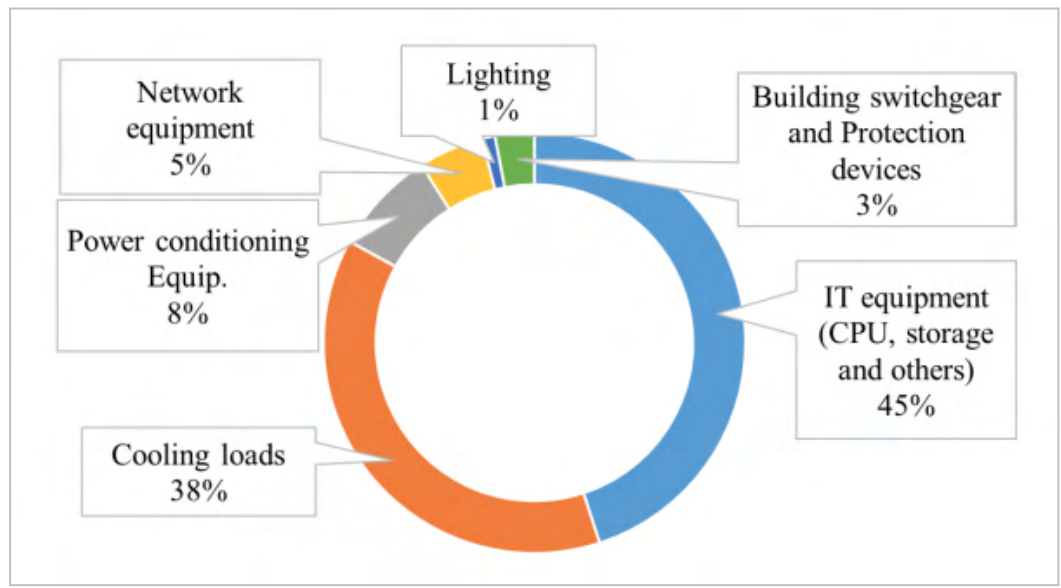

*Figure 12 - Power consumption proportionality according to Ahmed et al., 2021*

Water

Water plays a crucial role in certain data center cooling systems that depend on evaporation, which presents one of the most common cooling solutions. Most data centers source water from public supplies and sizable facilities often have additional nearby water bodies to draw from in case of a shortage.

On average, evaporative cooling systems use approximately 1,800 to 2,900 liters of water per megawatt per hour (*Water Technology, 2021*; *Mytton, 2021*). However, this figure greatly depends on the efficiency of the cooling system, as well as seasonal variations and the location of the data center.

To put these numbers into perspective, an average U.S. family uses about 1,136 liters of water per day, meaning that a large data center uses roughly the same amount of water as 720 U.S. families. Meanwhile, a golf course uses around 1,000 liters of water per hour, while a coal power plant evaporates up to 90,000 liters per hour per megawatt. Therefore, a recent study found that in the US, the water used for cooling in data centers is only a fraction—less than 1%—of the water used to generate the electricity that powers it (*Data Center Dynamics, 2021a*).

### Skilled personnel

Operating a data center requires constant infrastructure maintenance and the replacement of parts and, thus, a small number of trained professionals. E.g., [Meta reports](#) about 100 permanent jobs to run a large data center

## Costs

I provide estimates for the cost of different data center inputs. This section should be read with caution as there are few detailed breakdowns publicly available.

### Construction costs

#### Supporting infrastructure

The construction of supporting infrastructure for a data center typically costs **$5M- $15M per MW** of power consumption (this is [excluding hardware](#)) ([*RBC, 2021*](#) page 105; [*Data Centre Dynamics, 2021b*](#); [*Dgtl Infra, 2022b*](#)). The final cost depends on the desired level of uptime and security as well as on the location. E.g., construction is cheapest in Chinese and Indian cities and most expensive in highly developed metropolitan areas such as Tokyo, Zurich, and the Bay Area ([*Statista, 2022h*](#)).

Considering that a typical new data center has a capacity of about 20 MW (according to the draft report), typical investments range from $100M to $200M.
**Crypto mining**[43] data centers have the lowest requirements for uptime and security and thus only cost a fraction of a typical data center. Industry experts reported costs as low as $500k per MW. Meanwhile, the largest data center projects cost more than a billion dollars, such as Meta's [Altoona Data Center](#)[44].

Additional useful indicators of the cost to set up a data center are some of the largest supercomputers in the [TOP500 list](#) (Table 2). Typical investments are between $10M and $30M per MW, though note that this includes hardware. Since supercomputers often rely on customized hardware, such projects are likely considerably more expensive than a typical data center. Furthermore, large cloud providers have developed considerable economies of scale, enabling them to cut costs for their data centers.

---

[43] Crypto mining is the process where transactions are verified and added to a blockchain ledger by using computational power. Miners earn cryptocurrency tokens as a reward.
[44] Meta does not specify whether this includes hardware. Furthermore, an industry expert cautioned against taking such numbers literally, as Big Tech companies may face incentives to underreport their investment.

| Name | Location | Power capacity [MW] | Investment [million USD] | Investment per MW [million USD] |
|---|---|---|---|---|
| **Aurora** | US, IL  9700 … | 60 | 500 | 8.3 |
| **Frontier** | US, TN  Oak R… | 40 | 600 | 15.0 |
| **Fugaku** | Japan  RIKE… | 40 | 1,000 | 25.0 |
| **LUMI** | Finland  CSC-… | 6 | 211 | 35.2 |
| **Leonardo** | Italy  Tecno… | 9 | 262 | 29.1 |
| **Summit** | US, TN  Oak R… | 13 | 200 | 15.3 |

*Table 2: Power capacity and investment of selected supercomputers*

Hardware

NVIDIA's DGX H100 system, containing eight H100 GPUs, currently costs around $325k and has a power capacity of 10.2 kW (*DELTA computer, 2023*). For a Fermi estimate, assume that a data center's hardware is based entirely of DGX systems[45]. This implies 1MW of IT power capacity would consist of about 800 GPUs[46] and cost around $32M.[47] Note that this estimate is imprecise as the cost of the DGX system likely changes when ordering large quantities[48].

Hardware for a typical data center, not optimized for AI, is considerably cheaper.

Construction time

Large-scale data center projects are typically completed in under a year if best practices are applied (*Stack Infrastructure, 2021, draft report, industry experts*). It is common practice to use a phased approach, meaning parts of the data center become operational while other sections are still being built. This strategy allows for a quicker return on investment (*Stack Infrastructure, 2021*).

Overall, the construction speed can be influenced by various factors. Using pre-fabricated, modular components and experienced contractors significantly accelerates construction. A lower uptime requirement, implying fewer redundancy components and less security, also speeds up the process. Early supply chain coordination and fast permitting processes, especially in less regulated jurisdictions, further increase efficiency. Lastly, proximity to existing infrastructure can reduce the need for additional construction, like power lines or connectivity cables, saving time (*draft report*).

---

[45] In addition to the DGX systems, an AI compute cluster would also require hardware for storage, management, fabric, and switches. However, these collectively constitute only 10% of the total power capacity and are considerably cheaper than the DGX systems (*NVIDIA, 2023*).
[46] 1 MW/ 10.2 kW * 8 GPUs = 784 GPUs
[47] 1 MW/ 10.2 kW * $325k = $31.9M
[48] Industry experts report contradicting information about the price changes for bulk orders of GPUs. Some claim large customers get considerable discounts whereas others report that cloud providers may even pay a larger amount of money to buy large quantities of GPUs.

Operating costs

Similar to construction costs, data center providers hardly ever publish detailed breakdowns for operating costs. However, some estimates can provide basic references.

One kWh in the US typically costs around $0.125 ([YCharts, 2023](#)). Assuming a utilization of 75%[49] and a PUE of 1.2, operating 1MW of hardware for one year costs around $1M[50] for electricity.

Meanwhile, water costs are considerably lower. The average price in the US is about $0.0025 per liter ([Statista, 2021](#))[51]. A data center relying on evaporative cooling consumes up to 2,900 liters of water per megawatt per hour ([Water Technology, 2021](#); [Mytton, 2021](#)). This accumulates to an annual water bill of $64k per MW[52].

[Meta](#) reports 300+ operational jobs for one of its data centers. Since the company added several more than 300 MW to the local power grid, the data center campus' power capacity is likely several hundred MW. This suggests between one and five operational jobs per MW. A data center operator earns about $55k annually ([glassdoor, 2023](#)), implying operating costs of $55k-$275k for personnel.

From conversations with experts, other considerable operating costs include property taxes and sales taxes and replacing and maintaining infrastructure and hardware.

([Back to summary](#))

---

[49] Utilizations are typically lower, e.g. in [this summary](#) of ML workloads, few exceed 60%.
[50] 1 MW * 8760 hours * 75% * 1.2 * $0.125 per kWh = $986k
[51] Industry water prices are likely well below household water prices.
[52] 8,760 hours * 2,900 liters per hour * $0.0025 per liter = $63,510

# 5) Potential limiting factors for rapid industry growth

AI could be a technology as transformative as the industrial revolution (*Dafoe, 2020*). This implies the possibility for rapid growth in demand for AI applications, necessitating a quick increase in global data center capacity. This section explores which of the features introduced in sections on key Inputs for data center construction and operation appear to be potential bottlenecks in such scenarios.

*Note: This part of the report is considerably more speculative, and the given estimates are particularly subjective.*

## Power

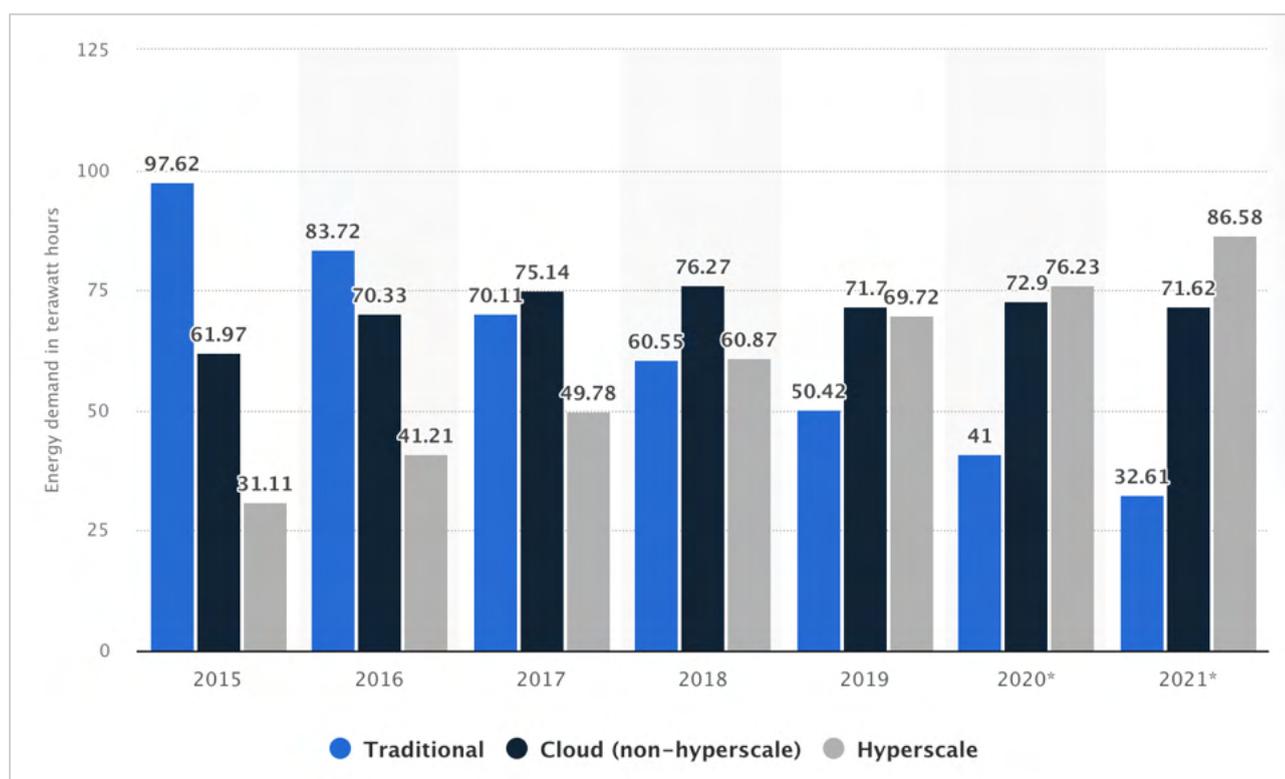

*Figure 13 - Power consumption of different industry sectors as reported by Statista, 2023c.*

Globally, data centers already consume around 1 - 2 % of global electricity[53][54] (*IEA, 2022*) and up to 18% of the local electricity in certain regions such as Ireland (*Euronews, 2023*). While power consumption is a major operating cost, it primarily poses a challenge to the industry's growth because free capacity in the power grid is increasingly hard to find (*industry experts, Data Center Frontier, 2023*). Major hubs, such as Ireland, Singapore, Amsterdam, and Virginia, are already unable to provide sufficient power for new projects, resulting in temporary pauses of data center construction (*Data Center Dynamics, 2022b*; *Data Center Dynamics, 2022c*; *Data Center Dynamics, 2021*; *Insider Intelligence, 2022*).

---

[53] ~45 GW of 2,6 TW (IEA, 2021)
[54] Note that about a third of this (15 GW) was used for crypto mining in 2021 (*IEA, 2022*; *For more on this sector, see Appendix: Crypto mining*).

Yet, despite considerable industry growth, efficiency improvements have largely kept global energy consumption constant over the past five years (Figure 13). However, it is unclear if efficiency increases will continue, with some sources arguing that even absent transformative growth, power consumption of the industry could rise considerably as efficiency increases abate (*Masanet et al., 2020*; *Statista, 2022i*).

Furthermore, internal analysis suggests that certain individual data center projects currently grow in energy consumption at an unsustainable pace.[55]

However, in scenarios of AI speeding up scientific progress, AI technology could contribute to making data centers more efficient. E.g., in 2016, DeepMind used ML to reduce power usage of the cooling systems in Google's data centers by 40%. Similar feedback could occur if AI were used in hardware or data center design.

## Cooling

As Moore's law is slowing, hardware becomes increasingly energy-dense (*industry expert, Techspot, 2022*). E.g., NVIDIA's most recent GPU architectures, the V100, A100, and H100, had a peak power consumption of 300 W, 400 W, and 700 W each. Besides an increase in power consumption, this also necessitates more effective cooling systems. Should energy density continue to increase at the current pace, there may be considerable engineering challenges to cooling future hardware, and costs for data centers could increase. Further research is required to assess the likelihood of such a scenario. Industry experts are so far optimistic that new cooling paradigms, such as liquid cooling, will be able to provide efficient cooling.

## Hardware

Due to the complexity of the semiconductor supply chain, computer hardware itself will likely be the most severe bottleneck for scaling compute capacity in scenarios of transformative growth. In fact, NVIDIA already struggles to keep up with GPU demand (*digitaltrends, 2023*).

Besides chips, some industry experts predict that connectivity equipment, relying on sophisticated chips as well as fiber optics technology, could similarly be difficult to scale up in the future.

However, assessing bottlenecks in the semiconductor supply chain remains outside the scope of this report. For an analysis of its current state, see *Khan et al., 2021*.

## Location/ suitable sites

According to industry experts, Big Tech companies are increasingly having difficulties finding suitable sites for large data centers (>100 MW). This is due to the combination of several requirements, most notably a reliable and cheap power supply (as mentioned above), and high bandwidth connectivity. Additionally, regulatory issues appear to be a burden, most recently primarily related to sustainability requirements (inudstry expert). Further factors limiting the available locations include water supply and skilled staff availability.

---

[55] For more information, please reach out to mail@konstantinpilz.com.

In fact, suitable sites seem to be so difficult to acquire in urban areas that data center providers increasingly construct multi-story facilities—despite construction and cooling being considerably more expensive compared to single-story data centers (*Cushman & Wakefield, 2022*).

## Supply chain issues

As of early 2023, many data center inputs still had longer waiting times than usual, mostly because of demand for data centers spiking during the pandemic and due to supply chain issues in connection with ongoing lockdowns in China. This meant that many materials were delayed by dozens of weeks, e.g., aluminum 40 weeks, PVC 20 weeks, and electrical equipment 40-50 weeks (*Turton Bond, 2022*). Illustratively, in 2022, only 23% of data center suppliers reported reliably being able to deliver on time (*Davis et al., 2022*).

Interestingly, experts reported that rather than specific materials presenting the most prominent bottleneck, it was more the high variety of inputs required that slowed the construction of data center projects.[56] However, industry analysts and equipment suppliers at an industry conference in March 2023 perceived these issues as temporary and expected them to be resolved quickly.

Nevertheless, the COVID-19 pandemic presents a meaningful analogy to scenarios of rapid growth in transformative AI scenarios, as it involved a) a sudden spike in data center demand[57] and b) considerable disruptions in global trade, which could also occur as AI worsens diplomatic relationships between countries.

The fact that the industry experienced considerable supply chain issues indicates that infrastructure shipments would pose a limit to how quickly data center capacity could be scaled up. However, even before equipment supply becomes a severely limiting factor, other bottlenecks, such as semiconductor manufacturing and available power capacity, would likely be a larger issue. Further research is required to disentangle such scenarios.

## Minor bottlenecks

The following issues appear unlikely to be considerable bottlenecks but are frequently discussed in the industry.

### Skilled personnel

The industry currently reports a shortage of skilled IT professionals, with more than half of respondents in a recent survey reporting difficulties with finding qualified candidates (*Tee, 2021*). This situation is exacerbated by the fact that many data center professionals are going to retire in the coming years (*Uptime Institute, 2023*).[58]

---

[56] Yet, certain materials such as batteries and power generators were repeatedly mentioned to be particularly delayed.
[57] Due to an increase in the use of video communication and streaming services.
[58] Interestingly, about 19% of respondents in another survey think AI systems will replace parts of their operations staff in the next five years, and an additional 52% think it will happen in more than five years, indicating that at least part of the problem will be solved through automation (*Davis et al., 2022*).

Water

While water availability limits the potential locations of data centers to some extent ([NBC news, 2021](#)), it is insignificant for the industry at large compared to other location factors such as power supply, connectivity, and regulatory burden. (Also see the section on [Location trends](#))

Conclusion

Due to the mentioned limiting factors of spare power capacity and supply chain logistics, it seems unlikely that the data center industry could grow faster than 40% per year, even in scenarios of explosive demand for AI. However, the semiconductor supply chain may well present a bottleneck even before such growth can be realized. Models for forecasting future AI capacity should take the mentioned bottlenecks into account.

([Back to summary](#))

# 6) Acknowledgements

Thank you to Lennart Heim for mentoring me on this project. He guided me from the beginning, scoped the questions for this project, helped me prioritize what to investigate, and provided in-depth feedback several times. Thank you to Yannick, Peter, Sam, Max, and Ben for feedback on different versions of this report. Thank you to various industry experts who answered my questions and provided commentary.

# 7) Appendix

## Appendix: How many data centers have a capacity of > 100 MW?

- Anchors
  - My general impression from expert consultations and looking at a variety of individual projects is that large tech companies' data centers are predominantly larger than 100 MW. For anecdotal evidence, see these reports: gtm, 2020; Volt Energy, 2022; Computerworld, 2017[59].
  - As a lower bound, Meta reports 17 data centers in the US, and they report having added 6,400 MW to the power grid.
    - Assuming a capacity factor of 33% for renewables, this indicates a capacity of 125 MW per data center.[60] However, Meta likely procures the majority of its electricity from already existing infrastructure. Therefore, this can only provide a lower bound.
  - Google reports 24 data centers, which are likely of a similar size.
  - AWS reports several classes of data centers in a compliance certificate, including 29 labeled as "Data center location," which are likely the largest ones.
  - Microsoft reports 14 data centers labeled as "Microsoft data center" which are likely the largest builds.
  - Overall, it seems reasonable that at least 80% of them have a power capacity of more than 100 MW. This provides a total of 67 data centers so far.
  - I assume that China has about 50% as many data centers >100 MW as the large US cloud providers. This adds up to a total of around 100 large data centers.
  - Other tech companies such as Apple and Oracle likely similarly have large data centers.
  - I found further scattered reports of other companies and colocation facilities reaching a similar max power load.
- Overall, this gives an estimate of 140 data centers (110 - 225, 70% confidence interval).
- Sanity check: Global data center electricity consumption is about 45,000 MW (See section on Power)
  - However, this is the *actual* power consumption, not the max load of the data centers. Data centers typically only utilize a fraction of the total power consumption (*GREENEX DC, 2022*).
  - It seems unlikely that > 100 MW data centers account for more than 50% of this → < 22,5000 MW from them.
  - The average >100 MW data center is probably about 200 MW.
  - Assuming that they operate at 50% of their max capacity, this means there are probably fewer than 225 data centers with a max capacity of > 100 MW.[61]

(Back to summary)

---

[59] I expect each reports only a fraction of the power capacity of the data center in question, though it provides a lower bound.
[60] 6,400 * 0.33 / 17 = 124.2
[61] 22,500 / (200 * 0.5) = 225

## Appendix: Crypto mining

Crypto mining is a significant share of the data center industry, accounting for about a third of its global energy consumption (*Cambridge Centre for Alternative Finance, 2023*). Bitcoin, by far, consumes the most energy, likely about two-thirds. Ethereum consumes most of the remaining third (*Moneysupermarket, 2021*). Crypto mining uses GPUs or specialized hardware such as ASICs. While small setups are possible, it seems that most mining takes place in designated data centers, and some companies have even specialized in these, e.g., RWB.

While China used to be the most important actor, accounting for as much as 75% of the global hash rate (mining rate), it looks like the government banned the use and mining of cryptocurrencies in 2021, greatly reducing the mining rate. (*World Economic Forum, 2022*; *BBC, 2021*) This is interesting information as it indicates that China, at some point, used electricity of ~5GW for mining, more than 10% of today's global data center energy consumption. Crypto is now increasingly mined where energy is cheap, e.g., in Kazakhstan, powered by coal. The US and China are the other major countries. (*Statista, 2023b*)

(Back to summary)